\documentclass[reprint,superscriptaddress,amsmath,amsfonts,amssymb,aps,prl,showkeys]{revtex4-1}
\usepackage{graphicx}
\usepackage{mathtools}
\usepackage{bm}
\usepackage[makeroom]{cancel}
\usepackage{amssymb}
\usepackage{dcolumn}
\usepackage{fullpage}
\usepackage{amsmath}
\usepackage{multirow}
\usepackage{physics}
\usepackage{dcolumn}
\usepackage{bm}
\usepackage{xcolor}
\usepackage{upgreek}

\begin{document}

\title{Nonlocal Spin Dynamics in the Crossover from Diffusive to Ballistic Transport}

\author{Marc Vila}
\affiliation{Catalan Institute of Nanoscience and Nanotechnology (ICN2), CSIC and BIST, Campus UAB, Bellaterra, 08193 Barcelona, Spain}
\affiliation{Department of Physics, Universitat Aut\`onoma de Barcelona, Campus UAB, Bellaterra, 08193 Barcelona, Spain}
\author{Jose H. Garcia}
\affiliation{Catalan Institute of Nanoscience and Nanotechnology (ICN2), CSIC and BIST, Campus UAB, Bellaterra, 08193 Barcelona, Spain}
\author{Aron W. Cummings}
\affiliation{Catalan Institute of Nanoscience and Nanotechnology (ICN2), CSIC and BIST, Campus UAB, Bellaterra, 08193 Barcelona, Spain}
\author{Stephen R. Power}
\affiliation{Catalan Institute of Nanoscience and Nanotechnology (ICN2), CSIC and BIST, Campus UAB, Bellaterra, 08193 Barcelona, Spain}
\affiliation{Universitat Aut\`onoma de Barcelona, Campus UAB, Bellaterra, 08193 Barcelona, Spain}
\affiliation{School of Physics, Trinity College Dublin, Dublin 2, Ireland}
\author{Christoph W. Groth}
\affiliation{Univ. Grenoble Alpes, CEA, IRIG-PHELIQS, 38000 Grenoble, France}
\author{Xavier Waintal}
\affiliation{Univ. Grenoble Alpes, CEA, IRIG-PHELIQS, 38000 Grenoble, France}
\author{Stephan Roche}
\affiliation{Catalan Institute of Nanoscience and Nanotechnology (ICN2), CSIC and BIST,
Campus UAB, Bellaterra, 08193 Barcelona, Spain}
\affiliation{ICREA--Instituci\'o Catalana de Recerca i Estudis Avan\c{c}ats, 08010 Barcelona, Spain}

\begin{abstract}
Improved fabrication techniques have enabled the possibility of ballistic transport and unprecedented spin manipulation in ultraclean graphene devices. Spin transport in graphene is typically probed in a nonlocal spin valve and is analyzed using spin diffusion theory, but this theory is not necessarily applicable when charge transport becomes ballistic or when the spin diffusion length is exceptionally long. Here, we study these regimes by performing quantum simulations of graphene nonlocal spin valves. We find that conventional spin diffusion theory fails to capture the crossover to the ballistic regime as well as the limit of long spin diffusion length. We show that the latter can be described by an extension of the current theoretical framework. Finally, by covering the whole range of spin dynamics, our study opens a new perspective to predict and scrutinize spin transport in graphene and other two-dimensional material-based ultraclean devices.
\end{abstract}

\maketitle

Since the seminal work of Tombros and coworkers \cite{Tombros2007}, who first measured long spin diffusion length in graphene nonlocal spin devices, a considerable number of studies have explored how to improve the material quality and the efficiency of spin injection and detection so as to reach the upper limit of spin transport \cite{Tombros2008, Han2011, Dlubak2012, Zomer2012, Guimaraes2014, Han2014, Gurram2016, Kamalakar2015, Serrano2019, Gebeyehu2019}. After fifteen years of progress, the fabrication of ultraclean (ballistic) graphene devices is now a reality with mean free paths reaching hundreds of nanometers and as long as tens of $\upmu$m at lower temperatures \cite{Wang2013dean, Banszerus2015nano}. Theoretical analysis of experimental data is usually based on the spin diffusion equations \cite{Fabian2007, Maassen2012, Wojtaszek2014, Sosenko2014, Idzuchi2014, Idzuchi2015, Obrien2016}, but their validity in new regimes of spin transport, especially the ballistic regime, is under question \cite{Roche2015, Gurram_2018}. A solution is to use quantum transport simulations to describe the spin dynamics in a realistic device geometry \cite{Tang2000, Zainuddin2011}. However, theoretical effort in this direction is currently lacking, which not only limits the understanding of spin transport in ultraclean devices but also restrains further improvement for spintronic applications based on two-dimensional materials and van der Waals heterostructures \cite{Lin2019}.

In this Letter we use quantum transport simulations to explore the physics of spin dynamics in graphene nonlocal spin valves (NSVs). By calculating exactly the nonlocal resistance $R_\text{nl}$, we are able to capture regimes that conventional analysis fails to describe. In the diffusive regime we show that the typically overlooked drain and reference electrodes (see Fig.~\ref{fig_F1} below) play a fundamental role in limiting $R_\text{nl}$ when spin relaxation is weak. When approaching the quasiballistic regime of spin transport, simulated Hanle precession curves reveal the failure of the diffusive equations. By extending the theory of spin diffusion following Refs.~[\citenum{Jedema2002, Petitjean2012}], we obtain more general formulas for properly obtaining the spin diffusion length $\lambda_\text{s}$ in the former case, and for understanding the evolution of Hanle curves in the crossover from diffusive to ballistic transport. Our findings demonstrate that brute force quantum simulation of nonlocal transport is fundamental to properly analyze spin dynamics in unconventional regimes and for allowing a direct comparison with experimental data.

Lateral NSVs are widely used to probe spin transport in disordered materials because they decouple electrical currents from spin currents, allowing for better device sensitivity \cite{Johnson1985, Johnson1988, Jedema2002, Zutic2004, Fabian2007, Wu2010}. In such a device configuration, shown in Fig.~\ref{fig_F1}, a ferromagnetic (FM) electrode (labeled ``2'') drives a spin-polarized current $I_0$ to a drain electrode (``1''), a spin accumulation develops below the FM, and this spin diffuses to the right along the channel. This spin is detected by another FM contact (``3'') as a nonlocal voltage $V_\text{nl}$, which is normalized to a nonlocal resistance $R_\text{nl} \equiv V_\text{nl}/I_0$. Owing to inherent spin relaxation, $R_\text{nl}$ usually decays exponentionally with channel length $d$, $R_\text{nl} \propto \exp(-d/\lambda_\text{s})$. Extracting $\lambda_\text{s}$ from the length dependence of $R_\text{nl}$ is quite difficult experimentally, but this can be avoided by applying a perpendicular magnetic field which provokes precession and additional dephasing of the spins. According to traditional spin diffusion theory, $R_\text{nl}$ then takes the form \cite{Jedema2002, Takahashi2003, Zutic2004, Fabian2007, Takahashi2008, Wu2010}
\begin{equation}\label{eq_normalHanle}
R_\text{nl} = \frac{P_\text{i} P_\text{d}}{2 w \sigma}\text{Re}\left\{\frac{e^{-d \alpha}}{\alpha}\right\},
\end{equation}
where $\alpha=\sqrt{\frac{1}{\lambda_\text{s}^2}+ i \frac{\omega}{D}}$, $P_\text{i}$ ($P_\text{d}$) is the polarization of the injector (detector) FM contact, $\sigma$ is the electrical conductivity, $w$ is the channel width, $D$ is the diffusion coefficient, and $\omega=g\mu_\text{B}B/\hbar$ is the Larmor spin precession frequency induced by the magnetic field $B$ with $g$ the g-factor, $\mu_\text{B}$ the Bohr magneton, and $\hbar$ the Planck constant. By fitting this expression to a measurement of $R_ \text{nl}$ vs.~$B$, one can extract the spin diffusion length $\lambda_\text{s}$.

This approach, known as a Hanle measurement, has been a cornerstone of the exploration of spin dynamics in a large variety of materials including metals \cite{Jedema2002, Villamor2015}, semiconductors \cite{Lou2007, Fabian2007, Salis2010, Erve2015, Hamaya2018}, and graphene \cite{Han2014, Roche2015}. However, Eq.~(\ref{eq_normalHanle}) is based on several assumptions which may be violated in ultraclean devices. Specifically, Eq.~(\ref{eq_normalHanle}) assumes that transport is fully diffusive, that relaxation is fast enough so that no spin signal reaches the reference electrode (lead 4), and neglects diffusion along the left-hand part of the NSV (between leads 2 and 1). It is therefore important to revisit and extend the current theoretical framework to cope with new spin transport regimes which are emerging in today's ultraclean nonlocal spin devices.

\begin{figure}[htbp]
\centering
\includegraphics[width=\columnwidth]{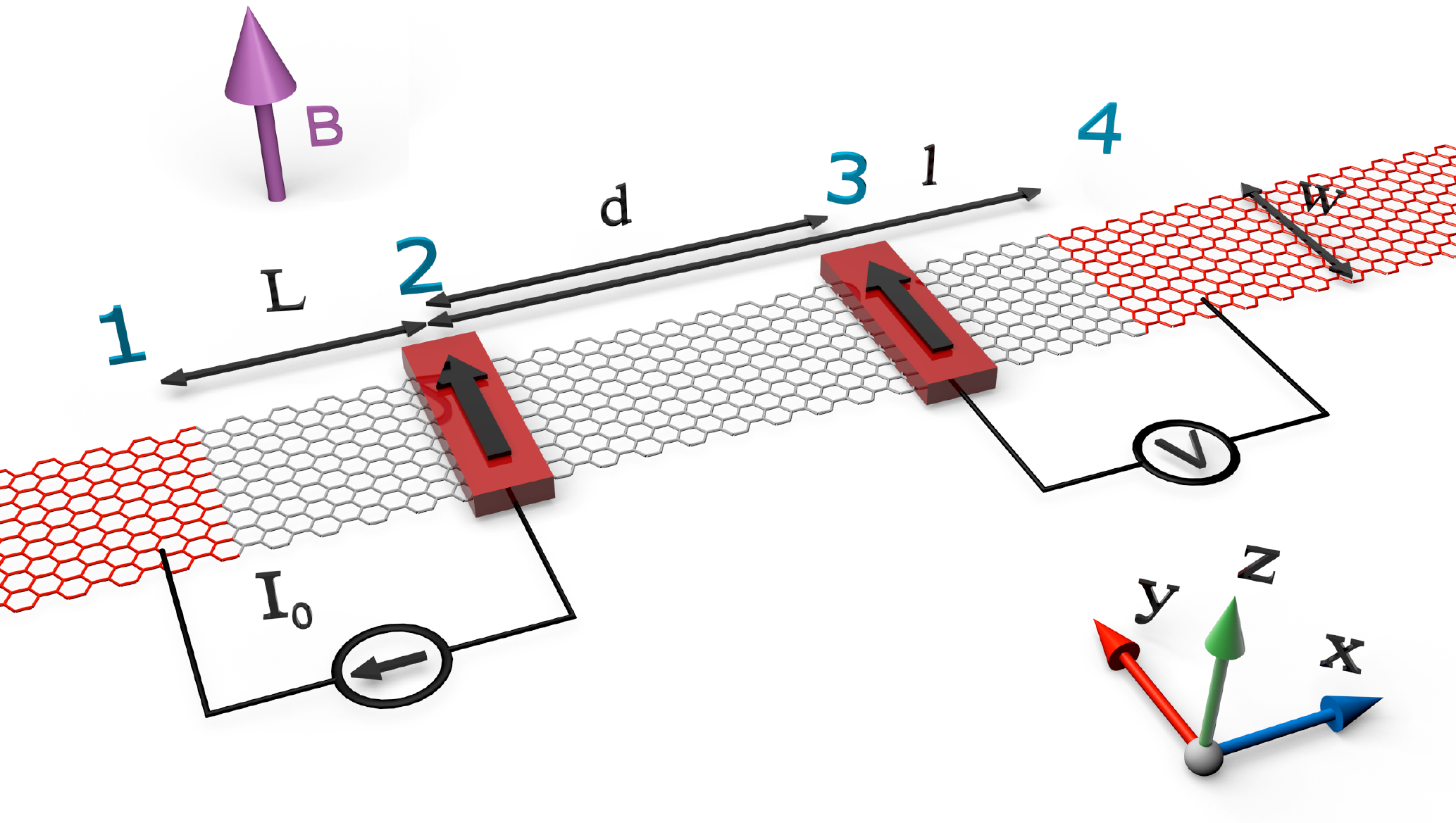}
\caption{Sketch of the lateral nonlocal spin valve. Red (black) regions denote the contacts (sample). The injector and detector contacts, labeled 2 and 3 respectively, are ferromagnetic with their magnetization indicated by arrows. Contacts 1 and 4 represent the drain and reference electrodes, respectively.}
\label{fig_F1}
\end{figure}

To study the behavior of graphene NSVs, we employ the Landauer-B\"{u}ttiker formalism, as implemented in Kwant \cite{Groth2014}, to the device setup in Fig.~\ref{fig_F1}. The graphene layer is described in a single-$\pi$-orbital tight-binding basis, with a Hamiltonian given by
\begin{align}\label{eq_Hamil}
\mathcal{\hat{H}}&= t \sum_{\langle i,j\rangle}  c_{i \eta}^\dagger c_{j \eta}   +  \sum_i \delta	U_i c_{i \eta}^\dagger c_{i \eta}   +  \sum_{i \eta\eta'} c^\dagger_{i \eta}   [\bm{s} \cdot \bm{J}_i]_{\eta\eta'} c_{i\eta'} \nonumber \\
&+ \mu_\text{B} \sum_{i \eta\eta'} c^\dagger_{i \eta}   [\bm{s} \cdot \bm{B}_i]_{\eta\eta'} c_{i\eta'},
\end{align}
where $c_{i\eta}^\dagger$ ($c_{i\eta}$) is the creation (annihilation) operator with spin $\eta$ on site $i$. The first term ($t=-2.6$ eV) denotes nearest-neighbor hopping in the graphene honeycomb lattice. The second term is Anderson disorder defined by a random potential uniformly distributed at each site $i$, with $\delta U_i \in [-U/2, U/2]$. The third term is magnetic disorder mainly affecting the spin dynamics. It is defined as a magnetic exchange coupling with strength $J$ and random orientation at each site $i$, $\bm{J}_i=J\left[\sin(\theta_i)\cos(\phi_i) , \sin(\theta_i)\sin(\phi_i) , \cos(\theta_i) \right]$, with $\theta$ and $\phi$ spherical angles and $\bm{s}$ the spin Pauli matrices. The last term is the Zeeman exchange induced by an external magnetic field $\bm{B}$ (note that orbital effects of the magnetic field are neglected). In general, $U$ is taken to be much larger than $J$, such that $U$ dictates the charge transport regime, whereas the spin relaxation is driven by $J$. The modeling of the leads is described in Supplemental Material \cite{Suppmat}.

The calculation of $R_\text{nl}$ is performed by evaluating all transmission probabilities between different leads. We then construct the conductance matrix $G$ \cite{Datta1997} and solve the linear system $I=G V$, where $I$ and $V$ are vectors including the current and voltage conditions at each electrode. We fix a current $I_0$ from lead 2 (injector) to lead 1 (drain) while enforcing that no current flows in leads 3 (detector) and 4 (reference). This ensures zero charge current in the channel since any current going to the right from the injector will be compensated with an oppositely spin-polarized current injected by the reference lead. We also ground the drain ($V_1=0$) and solve the system to obtain the other voltages. The nonlocal resistance is then calculated as $R_\text{nl}=(V_3-V_4)/I_0$.

\begin{figure}[t]
\centering
\includegraphics[width=\columnwidth]{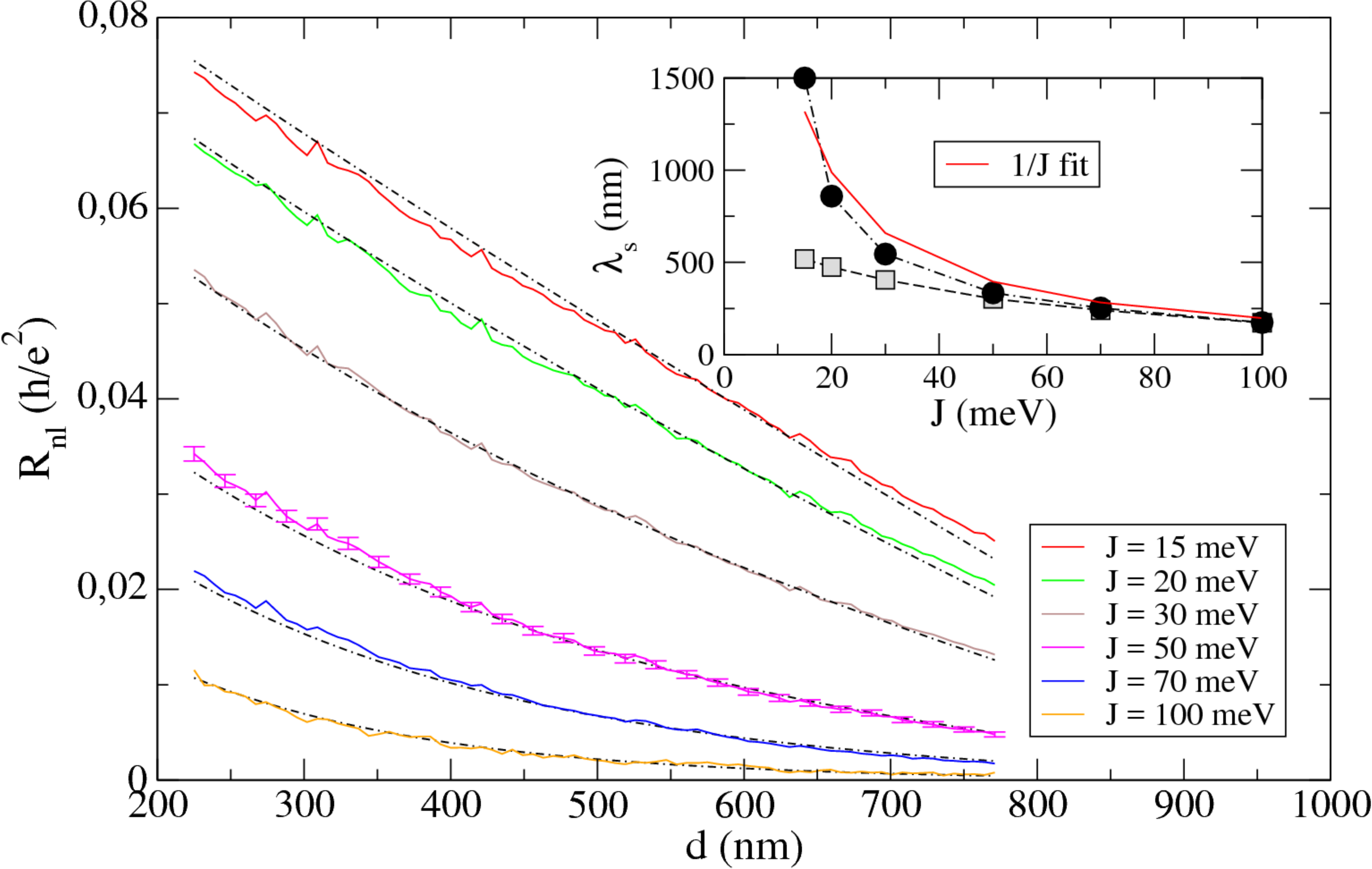}
\caption{$R_\text{nl}$ as a function of injector-detector distance for different strengths of magnetic disorder, with $l_\text{e} = 117$ nm. Error bars result from the averaging of several disorder configurations ($>130$). All curves have similar error bars. Black dot-dashed lines are the fits using Eq.~(\ref{eq_fullHanle}). Inset: comparison of $\lambda_\text{s}$ extracted from Eq.~(\ref{eq_normalHanle}) (gray squares) and Eq.~(\ref{eq_fullHanle}) (black circles). The red line indicates $1/J$ scaling of $\lambda_\text{s}$.}
\label{fig_F2}
\end{figure}

We first investigate spin dynamics in the diffusive regime of charge transport, which is identified from the scaling of the two-terminal conductance $G_\text{2T}$ with the channel length $x=L+l$ by removing leads 2 and 3. We evaluate the mean free path $l_\text{e}$ by fitting the numerical result to $G_\text{2T}=2e^2/h \times M l_\text{e}/x$, with $M$ the number of propagating modes per spin. Also, for $x \gg l_e$ we calculate the localization length $l_\text{loc}$ using $\langle \ln(G_\text{2T}) \rangle \propto -x/l_\text{loc}$. By choosing $w=20.1$ nm (164-aGNR), Fermi energy $E_\text{F}=0.4$ eV, $M=9$ and $U=1.04$ eV, we obtain $l_\text{e}=117$ nm and $l_\text{loc}=880$ nm \cite{Suppmat}. We take $L=250$ nm and $l=1000$ nm so that most of the transport occurs between $l_\text{e}$ and $l_\text{loc}$, and we compute $R_\text{nl}$ vs.~the channel length $d$ at $B=0$ for different magnetic disorder strengths $J$. The results are plotted in Fig.~\ref{fig_F2}.

For large values of $J$, $R_\text{nl}$ decays exponentially with channel length, as predicted by Eq.~(\ref{eq_normalHanle}). However, as spin relaxation slows with decreasing $J$, the decay of $R_\text{nl}$ becomes linear instead of exponential. Even for $J=0$, corresponding to $\lambda_\text{s} \rightarrow \infty$, there is a loss of spin signal with channel length \cite{Suppmat}, a result not captured by Eq.~(\ref{eq_normalHanle}). Conventional spin diffusion theory assumes the spin accumulation vanishes at $x\rightarrow + \infty$, or at least at $x=l$ \cite{Fabian2007, Wu2010}. However, this condition is violated for the lowest values of $J$ in our simulations, and may also be the case in recent experiments for which $\lambda_\text{s}$ reaches tens of $\upmu$m \cite{Drogeler2016spin}.

To describe the proper length dependence of $R_\text{nl}$, we solve the spin diffusion equations taking the full device geometry into account \cite{Suppmat}; not only are spins injected from lead 2, but leads 1 and 4 are explicitly included (lead 3 does not perturb the system). From this, $R_\text{nl}$ becomes
\onecolumngrid
\begin{equation}
R_\text{nl} = \frac{P_\text{i} P_\text{d}}{2 w \sigma } \text{Re} \left\{\frac{[\beta \cosh(L \alpha)+4 \sinh(L \alpha)] \cdot [\beta \cosh(\alpha(d-l)) - 4 \sinh(\alpha (d-l))]}{\alpha [4 \beta \cosh((L+l)\alpha)+(8+\beta^2/2)\sinh((L+l)\alpha)]} \right\},
\label{eq_fullHanle}
\end{equation}
\twocolumngrid
\noindent where $\beta=R_\text{c} w \sigma \alpha$ and $R_\text{c}$ is the contact resistance between leads 1 and 4 and the graphene. In the case of perfectly transparent contacts, the interface resistance is not zero but dictated by the  Sharvin resistance $R_\text{S} = h/(2 e^2 M)$ \cite{Rychkov2009, Borlenghi2011}. If one takes the limits $\lambda_\text{s} \ll L, l$, Eq.~(\ref{eq_normalHanle}) is recovered. Importantly, Eq.~(\ref{eq_fullHanle}) becomes linear when $\lambda_\text{s} \rightarrow \infty$,
\begin{equation}\label{eq_infiniteHanle}
R_\text{nl}=\frac{P_\text{i} P_\text{d}}{2w \sigma}\frac{(4R_\text{L}+R_\text{c})(-4 d + 4 l + R_\text{c} w \sigma)}{8R_\text{L} + 8R_\text{l} + 4R_\text{c}},
\end{equation}
where $R_\text{L}=L/w \sigma$ and $R_\text{l}=l/w \sigma$ are the sheet resistance of the left and right device regions, respectively. The black dot-dashed lines in Fig.~\ref{fig_F2} show the fits of the numerical results to Eq.~(\ref{eq_fullHanle}), indicating that this expression is able to capture the scaling of $R_\text{nl}$ for any value of $J$.

Equation (\ref{eq_infiniteHanle}) shows that when $\lambda_\text{s} \geq L, l$ the nonlocal spin signal still decays with length. This decay is no longer related solely to spin relaxation but also to charge diffusion and the presence of the leads. Recall that $R_\text{nl}$ depends on the conductance matrix $G$, which consists of the transmission between all leads and the imposed current/voltage conditions. In the limit of long $\lambda_\text{s}$, the drain and reference electrodes act as spin sinks, fixing the value of $R_\text{nl}$ in order to meet the conditions $I=I_0$ and $I=0$ at leads 1 and 4, respectively. We note that this spin sinking effect occurs despite the absence of spin relaxation in the leads. Rather it is the result of these leads absorbing and reinjecting spin current under the imposed boundary conditions. This contrasts with the contact-induced spin dephasing discussed in some experiments \cite{Maassen2012,Idzuchi2015,Amamou2016}.  Equation (4) shows that at the reference electrode ($d=l$) $R_\text{nl}$ is proportional to $R_\text{c}$ to leading order. Thus, in the limit of weak spin relaxation a small $R_\text{c}$ will suppress the nonlocal spin signal. Another consequence of long $\lambda_\text{s}$ is that the transmission between the drain and reference electrodes becomes crucial. The condition of zero charge current in the channel forces the injection of spin-down current from lead 4 to 1 so that lead 2 can inject up-spins that diffuse towards lead 3. If lead 4 (1) is unable to inject (absorb) down-spins to (from) the system, up-spins will not be able to diffuse along the channel and $R_\text{nl}$ will be suppressed. In Fig.~S3 \cite{Suppmat}, such effect is evidenced further by changing lead 1 from nonmagnetic to FM, which reduces $R_\text{nl}$ by more than three orders of magnitude. This suggests not employing FM materials for leads 1 and 4 in experiments. Another important consequence is that since Eq.~(\ref{eq_normalHanle}) does not account for this extra decay induced by leads 1 and 4, this reduction is absorbed in the value of $\lambda_\text{s}$, which will be therefore underestimated by Eq.~(\ref{eq_normalHanle}). This is shown in Fig.~\ref{fig_F2} (inset), where $\lambda_\text{s}$ is plotted vs.~spin relaxation strength. The gray squares are extracted from fits to Eq.~(\ref{eq_normalHanle}), while the black circles are from Eq.~(\ref{eq_fullHanle}). The spin diffusion length is the same when $\lambda_\text{s} < L, l$ (large $J$), but for small $J$ Eq.~(\ref{eq_normalHanle}) significantly underestimates the value of $\lambda_\text{s}$. According to the theory of spin relaxation arising from exchange fluctuations, $\lambda_\text{s}$ should scale as $1/J$ \cite{Fabian2007, Suppmat}, which is captured by the fits to Eq.~(\ref{eq_fullHanle}).

\begin{figure}[t]
\centering
\includegraphics[width=1 \columnwidth]{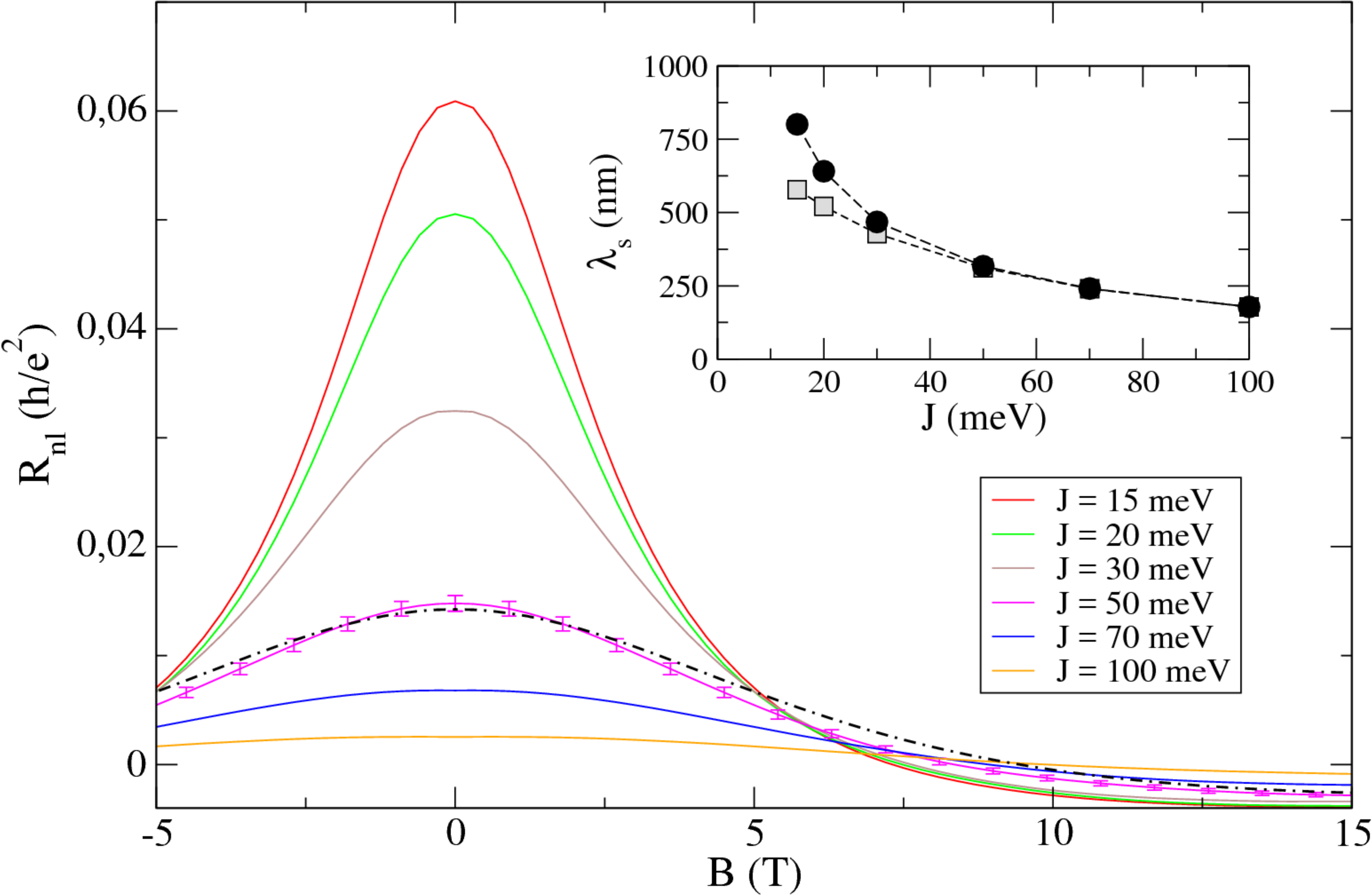}
\caption{Hanle spin precession curves for different strengths of magnetic disorder, with $l_\text{e} = 117$ nm and $d = 500$ nm. Error bars result from averaging several disorder configurations ($> 90$). All curves have similar error bars. Black dot-dashed line is the fit using Eq.~(\ref{eq_fullHanle}). Inset: comparison of $\lambda_\text{s}$ extracted from Eq.~(\ref{eq_normalHanle}) (gray squares) and Eq.~(\ref{eq_fullHanle}) (black circles).}
\label{fig_F3}
\end{figure}

We now extend the analysis to Hanle precession and plot $R_\text{nl}$ vs.~$B$ in Fig.~\ref{fig_F3} using a channel length of $d=500$ nm. We note that large magnetic fields are required for computational convenience (our device is smaller than in experiments) but no spurious effect is introduced since the Zeeman splitting remains much smaller than the subband energy separation and orbital effects are excluded. The simulation data are fitted with Eqs.~(\ref{eq_normalHanle}) and (\ref{eq_fullHanle}) using $D=v_\text{F}l_\text{e}$ and the resulting $\lambda_\text{s}$ are compared in Fig.~\ref{fig_F3} (inset). Similar to Fig.~\ref{fig_F2}, in the limit of weak spin relaxation $\lambda_\text{s}$ is underestimated when using the conventional equation.

To estimate how strong the underestimation of $\lambda_\text{s}$ may be in state-of-the-art devices, we calculate a Hanle curve with Eq.~(\ref{eq_fullHanle}) using realistic parameters ($L=5$ $\upmu$m, $l=20$ $\upmu$m, $d=15$ $\upmu$m, $\lambda_\text{s}=10$ $\upmu$m, $D=0.05$ m$^2$/s) and fit it with Eq.~(\ref{eq_normalHanle}). We obtain $\lambda_\text{s}=7.26$ $\upmu$m, about $25 \%$ less than the real value. More importantly, the spin lifetime ($\tau_\text{s}=\lambda_\text{s}^2/D$) is underestimated by nearly $100\%$; the real value is $2$ ns while the fit gives $1.05$ ns. These results thus call for a revised analysis of Hanle spin precession measurements taking into account the device geometry and our more general formula (Eq.~(\ref{eq_fullHanle})).

\begin{figure}[t]
\centering
\includegraphics[width=\columnwidth]{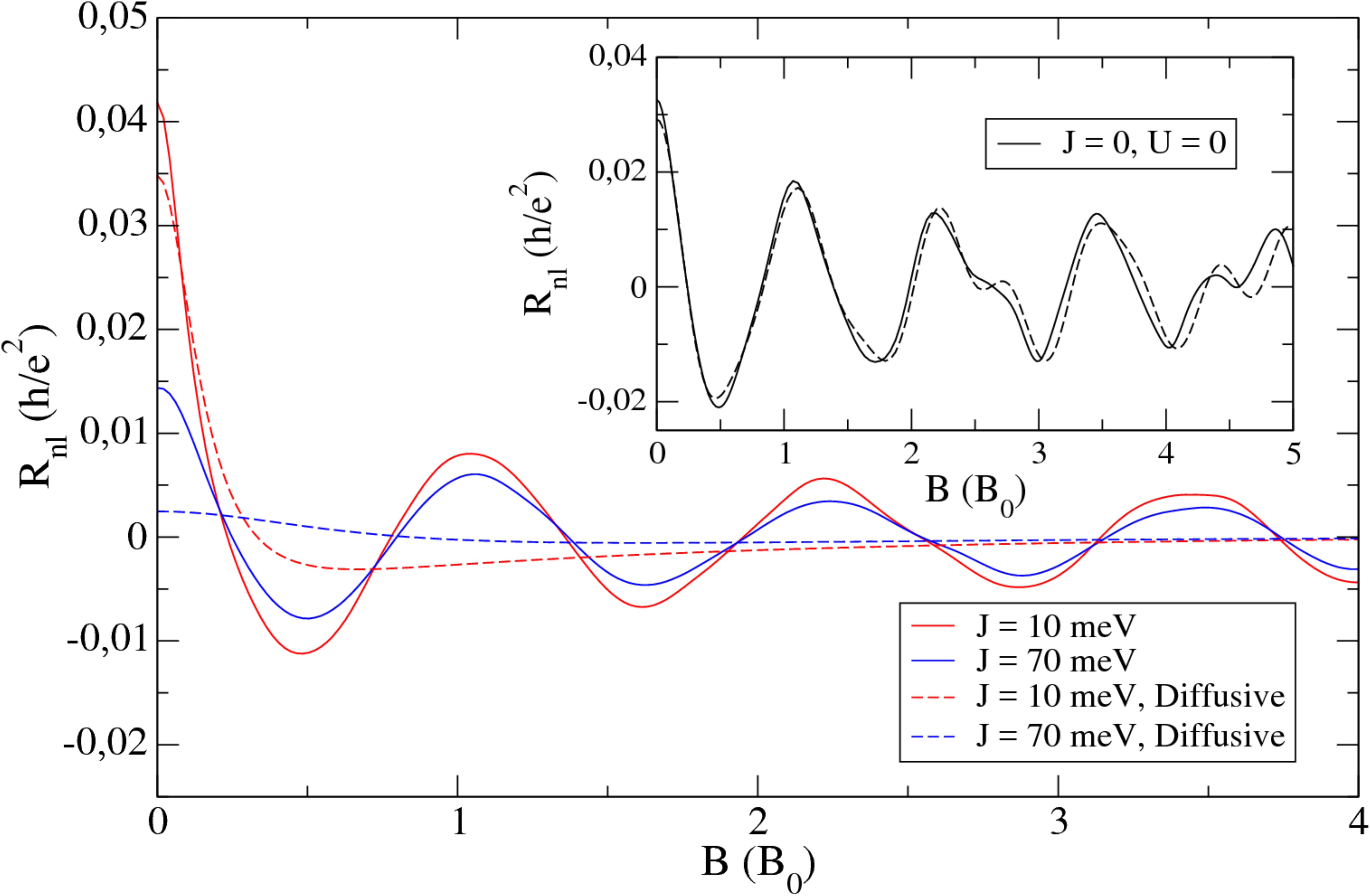}
\caption{Hanle spin precession curves in the quasiballistic regime, with $l_\text{e} = 487$ nm and $d = 500$ nm. Solid lines correspond to simulations (averaged from 12 disorder configurations), while dashed lines are fits using Eq.~(\ref{eq_fullHanle}). Inset: Case with $U=J=0$, solid (dashed) line shows the simulation ($R_\text{nl} \propto \sum_i^M  \cos\left(\gamma d B / v_{\text{F},i}\right)$).}
\label{fig_F4}
\end{figure}

Finally, we examine the quasiballistic limit. When $l_\text{e} \sim d$, only a few scattering events occur during transport through the channel. This situation has been discussed for spin relaxation in ultraclean graphene \cite{Cummings2016dephasing}, but little is known about its impact on Hanle measurements.  We keep all simulation parameters the same as before, including the channel length $d = 500$ nm, and reduce the Anderson disorder to $U=0.52$ eV, giving a mean free path $l_\text{e} \sim 500$ nm. The solid lines in Fig.~\ref{fig_F4} show the Hanle curves for this quasiballistic regime, and the dashed lines show fits using Eq.~(\ref{eq_fullHanle}). The behavior of $R_\text{nl}$ is now substantially different, especially with respect to the dependence on $B$. The unit $B_0$ corresponds to the magnetic field needed for spin to precess $2 \pi$ radians upon reaching the detector ($B_0=\frac{2\pi v_\text{F}^{\text{av}}}{\gamma d}$, with $\gamma = g \mu_\text{B}/\hbar$ the gyromagnetic ratio and $v_\text{F}^{\text{av}}$ the averaged Fermi velocity from all modes at the Fermi level). The first rotation of the spins occurs at $B_0=1$, but is followed by a dispersion of frequencies for larger $B$. This can be understood if we examine the origin of such oscillations. By performing a simulation in a purely ballistic regime, $U=J=0$, we observe in Fig.~\ref{fig_F4} (inset) that the main oscillation has the same period, but is superimposed with other frequencies. This arises because the nonlocal signal is the sum of each propagating mode moving at a different velocity. To verify this, we follow Ref.~\citenum{Jedema2002} and sum the contributions of spin over all transport times to get $R_\text{nl} \propto \sum_i^M  \cos\left(\gamma d B / v_{\text{F},i}\right)$ \cite{Suppmat}. By taking only the Fermi velocities of each mode of the system, the simulations are very well reproduced. 

From these results we can conclude that in the quasiballistic regime the scattering is weak enough for the precession to follow that of ballistic transport, but is also strong enough to average the beating pattern to one main frequency. This explains why neither Eq.~(\ref{eq_fullHanle}) nor the sum of cosines is able to fit the quasiballistic Hanle curves. This difficulty in capturing the crossover from diffusive to ballistic transport in a single expression highlights the importance of brute force quantum simulations to understand such a regime. Furthermore, in the supplemental material \cite{Suppmat} we simulate Hanle curves with different $l_\text{e}$ and find that when $l_\text{e} > d/3$, the spin dynamics enters the quasiballistic regime. Finally, in the limit of a 2D graphene flake, with most electrons moving at the same Fermi velocity, one would expect the signal to be determined by a single frequency. To observe this effect at low magnetic fields ($B \le 0.5$ T), the channel length needs to be $d \ge \frac{2\pi v_\text{F}}{\gamma B} \approx 50$ $\upmu$m. We highlight the fact that this analysis can be applied to other materials as well; depending on whether there are electrons moving at the same or different Fermi velocities, one can expect single or multiple precession frequencies, respectively.

In conclusion, we have performed fully quantum simulations that provides a more global picture of nonlocal spin transport when the material quality drives the system towards the quasiballistic regime, as well as an extended theoretical frame to analyze systems with long spin diffusion lengths. In this limit, the drain and reference electrodes become the limiting factors, and one should aim for these to be nonmagnetic and optimize their contact resistance to reach the upper limit for spin information transfer. Beyond guiding future nonlocal spin transport measurements in graphene devices, the developed methods and findings should be also relevant for other types of two-dimensional materials and van der Waals heterostructures.

\begin{acknowledgments}
We thank Ivan Vera-Marun for highly fruitful comments. M.V. acknowledges support from ``La Caixa'' Foundation. S.R.P. acknowledges funding from the Irish Research Council under the Laureate awards programme. X.W. acknowledges the ANR GRANSPORT funding.  All authors were supported by the European Union Horizon 2020 research and innovation programme under Grant Agreement No. 785219 (Graphene Flagship). ICN2 is funded by the CERCA Programme/Generalitat de Catalunya, and is supported by the Severo Ochoa program from Spanish MINECO (Grant No. SEV-2017-0706).
\end{acknowledgments}

%

\end{document}